\documentclass[12pt,a4paper]{article}
\usepackage{amsmath}
\usepackage[dvips]{graphicx}
\input epsf
\usepackage{times}
\begin{document}
\begin{titlepage}
\title{Elliptic flow in $pp$-collisions at the LHC}
\author{S.M. Troshin,
 N.E. Tyurin\\[1ex]
\small  \it Institute for High Energy Physics,\\
\small  \it Protvino, Moscow Region, 142281, Russia}
\normalsize
\date{}
\maketitle

\begin{abstract}
We consider collective effects in $pp$-collisions at the LHC energies related to presence
of the large orbital angular momentum in the initial state and role of this orbital momentum 
in the elliptic flow behavior. 
 \\[2ex]
\end{abstract}
\end{titlepage}
\setcounter{page}{2}

\section*{Introduction}
There were several collective effects in nuclear collisions  observed at RHIC and various experimental
observables relevant to their studies have been measured. In particular, there was observed 
ridge structure in the  near-side two-particle correlation function in the  secondary
particles production.
Similar structure in the two-particle correlation function was observed by
the CMS Collaboration \cite{ridgecms}. This is not an expected result.  Commonly,
 $pp$--collisions were  treated as  a kind of ``elementary''  ones or the reference
process for detecting deconfined phase formation in $AA$-collisions. 
It becomes evident that such a view should be  corrected.

The experimental observations of RHIC and LHC on the ridge in the two-particle correlation functions
 demonstrate that the hadronic matter is strongly correlated
and reveals high degrees of coherence when it is well beyond the critical values of density and temperature.
The other anisotropic flow measurements at RHIC also demonstrated such a collective behavior.
 
In this note  we discuss the elliptic flow in proton collisions on the base of approach \cite{intje} which 
was applied for discussion of the directed flow and the ridge in proton collisions \cite{ridge}. 
This  particular approach is based on the non-perturbative hadron dynamics. It should be noted that  the values of
 transverse momenta in the relevant experimental data are not too large. Despite that there are considerations of 
 the experimental results rendering to the perturbative QCD and parton radiation in the medium. 

\section{Elliptic flow in peripheral proton collisions}

We consider in this section peripheral hadronic collisions. These collisions suppose that the impact parameter
is different from zero. The mechanism of selecting such events in proton collisions will be considered
further for the LHC energies. Here we adopt  that impact parameter is non-zero and the reaction plane can
therefore be determined by the cumulant method.

There are several experimental probes  of collective dynamics. A  most widely discussed one
is the elliptic flow
\begin{equation}\label{v2}
v_2(p_\perp)\equiv \langle \cos(2\phi)\rangle_{p_\perp}=\langle \frac{p_x^2-p_y^2}{p_\perp^2}\rangle,
\end{equation}
which is the second Fourier moment of the azimuthal momentum distribution of the particles
 with a  fixed value of $p_\perp$.
The azimuthal angle $\phi$ is
an angle of the detected particle with respect to the reaction
plane, which is spanned by the collision axis $z$ and the impact parameter vector $\mathbf b$. The impact
parameter vector $\mathbf b$ is directed along the $x$ axis. Averaging is taken over large number
 of the events. Elliptic flow can be expressed  in covariant form in
 terms of the impact parameter and transverse momentum
 correlations as follows
 \begin{equation}\label{v2a}
v_2(p_\perp)=\langle \frac{(\hat{\mathbf  b}\cdot {\mathbf  p}_\perp)^2}{p_\perp^2}\rangle-
\langle\frac{(\hat{\mathbf  b}\times {\mathbf  p}_\perp)^2}{p_\perp^2}\rangle ,
\end{equation}
where $\hat{\mathbf  b}\equiv \mathbf  b /b$. 
To get some hints on the possible behavior of the elliptic flow in proton collisions,
it is useful to recollect what is known on this observable from nuclear collisions experiments.
Integrated elliptic flow $v_2$  at high energies
is positive and increases with $\sqrt{s_{NN}}$.
The differential elliptic flow $v_2(p_\perp)$ increases with $p_\perp$
at small values of transverse
momenta, then it becomes flatten in the region of the intermediate transverse
momenta and decreases at large $p_\perp$. 

It is also useful  to apply a simple geometrical ideas which imply existence of the elliptic
flow in hadronic reactions.
Geometrical notions for description
of multiparticle production in hadronic reactions were used by many authors, e.g  by Chou and Yang in \cite{chyn}.
In the peripheral hadronic collisions the overlap region has different sizes along the $x$ and $y$ directions.
According to the uncertainty principle we can estimate the value
of $p_x$ as $1/\Delta x$ and correspondingly $p_y\sim 1/\Delta y $
where $\Delta x$ and $\Delta y$ characterize the size of the
region where the particle originate from. Taking $\Delta x \sim
R_x$ and $\Delta y \sim R_y$, where $R_x$ and $R_y$ characterize
the sizes of the almond-like overlap region in transverse plane,
 we can easily
obtain proportionality of $v_2$ in collisions with fixed initial impact 
parameter to the eccentricity of the overlap
region, i.e.
\begin{equation}\label{exc}
v_2\sim \frac{R_y^2-R_x^2}{R_x^2+R_y^2}.
\end{equation}

The presence of correlations of impact parameter vector $\mathbf
b$ and $\mathbf p_\perp$ in hadron interactions follows also from the relation between
impact parameters in the multiparticle production:
\begin{equation}\label{bi}
{\mathbf b}=\sum_i x_i{ {\mathbf  b}_i}.
\end{equation}
Here  $x_i$ stand for Feynman $x_F$ of $i$-th particle, the impact
parameters ${\mathbf b}_i$ are conjugated to the transverse
momenta ${\mathbf p}_{i,\perp}$. 

The above  considerations are  based on the uncertainty principle and angular momentum
conservation, but they do not preclude the existence of the dynamical description, which will be discussed
in the next section

\section{Peripheral  inelastic proton-proton interactions at the LHC energies}

The particle
production mechanism  proposed in the model  \cite{intje,ptrans} takes into  account  the 
geometry  of the overlap region and dynamical properties of
the transient state in hadron interaction. This picture assumes  deconfinement at the initial stage of interaction.
The transient state here appears as a rotating medium of massive quarks and pions which hadronize and 
form multiparticle final state at the freeze-out stage.  Essential point for this 
rotation is the presence of a non-zero impact parameter in the collision.

In this section we consider  unitarity saturation as a dynamical mechanism leading to the  peripheral nature
 of inelastic collisions at the LHC energies. We also discuss the limiting energy dependence of the anisotropic
flows resulting from unitarity saturation in central collisons ( $b=0$).

In the untarization approach ($U$--matrix) the elastic scattering matrix in the impact
parameter representation has the form:
\begin{equation}
S(s,b)=\frac{1+iU(s,b)}{1-iU(s,b)}, \label{um}
\end{equation}
where $S(s,b)=1+2if(s,b)$ and $U(s,b)$ is the generalized reaction matrix, which is
considered to be an input dynamical quantity similar to the
eikonal function. Unitarity equation rewritten at high energies
for the elastic amplitude $f(s,b)$ has the form
\begin{equation}
\mbox{Im} f(s,b)=h_{el}(s,b)+ h_{inel}(s,b)\label{unt}
\end{equation}
where the inelastic overlap function
\[
h_{inel}(s,b)\equiv\frac{1}{4\pi}\frac{d\sigma_{inel}}{db^2}
\]
 is the sum of
all inelastic channel contributions.
Inelastic overlap function
is related to $U(s,b)$ according to Eqs. (\ref{um}) and (\ref{unt}) as follows
\begin{equation}
h_{inel}(s,b)=\frac{\mbox{Im} U(s,b)}{|1-iU(s,b)|^{2}}\label{uf},
\end{equation}
this function is the probability distribution over impact parameter of the inelastic interactions:
\begin{equation}\label{sinel}
\sigma_{inel}(s)=8\pi\int_0^\infty bdb \frac{\mbox{Im} U(s,b)}{|1-iU(s,b)|^{2}}.
\end{equation}
It should be noted that
\begin{equation}\label{imu}
\mbox{Im} U(s,b)=\sum_{n\geq 3} \bar U_n(s,b),
\end{equation}
where $\bar U_n(s,b)$ is a Fourier--Bessel transform of the function $\bar U_n(s,t)$, the form 
of this function can be found in 
\cite{ptrans}.

Since the form of the impact parameter dependence of the function function $h_{inel}(s,b)$ evolves from a 
central to the peripheral one
 at the LHC energies \cite{intja} and this function tends to zero at $b=0$ and $s\to\infty$, the mean multiplicity 
\[
 \langle n\rangle (s)=\frac{\int_0^\infty bdb  \langle n\rangle (s,b) h_{inel}(s,b)}
{\int_0^\infty bdb h_{inel}(s,b)}
\]
obtains a main input from the collisions with non-zero impact parameters.
One can suggest therefore that the events with average and higher multiplicity at the LHC energy $\sqrt{s}=7$ TeV
  correspond
to the peripheral hadron collisions \cite{intja}. Thus, at the LHC energy $\sqrt{s}=7$ TeV there is a dynamical
selection of the peripheral region in impact parameter space responsible for the inelastic processes. 
In the nuclear reactions
similar selection of the peripheral collisions is provided by the relevant  experimental adjustments. 

Of course, the standard inclusive cross-section for unpolarized particles
being integrated over impact parameter $\mathbf b $,  does not depend on the
azimuthal angle of the detected
particle  transverse momentum. It can be written with account for $s$--channel unitarity
in the following form
\begin{equation}
\frac{d\sigma}{d\omega}= 8\pi\int_0^\infty
bdb\frac{I(s,b,\omega)}{|1-iU(s,b)|^2}\label{unp}.
\end{equation}

When the impact parameter vector $ \mathbf {b}$ and transverse momentum ${\mathbf  p}_\perp $
of the detected particle are fixed
the function $I$ does
  depend on the azimuthal angle $\phi$ between
 vectors $ \mathbf b$ and ${\mathbf  p}_\perp $.
It should be noted that the impact parameter
$ \mathbf {b}$ is the  variable conjugated to the transferred momentum
$ \mathbf {q}\equiv \mathbf {p}'_a-\mathbf {p}_a$ between two incident channels
 which describe production processes
of the same final multiparticle state.
The dependence on the azimuthal angle $\phi$ can be written in explicit form through the Fourier
series expansion
\begin{equation}\label{fr}
I(s,\mathbf b, y, {\mathbf  p}_\perp)=\frac{1}{2\pi}I_0(s,b,y,p_\perp)[1+
\sum_{n=1}^\infty 2\bar v_n(s,b,y,p_\perp)\cos n\phi].
\end{equation}
The function $I_0(s,b,\xi)$ satisfies  to the
following sum rule
\begin{equation}\label{sumrule}
\int I_0(s,b,y,p_\perp) p_\perp d p_\perp dy=\langle n \rangle (s,b)\mbox{Im} U(s,b),
\end{equation}
where $\bar n(s,b)$ is the mean multiplicity depending on impact parameter.
Thus, the bare anisotropic  flow $\bar v_n(s,b,y,p_\perp)$ is related to the
measured  flow $v_n$  as follows
\[
v_n(s,b,y,p_\perp)=w(s,b)\bar v_n(s,b,y,p_\perp).
\]
where the function $w(s,b)$ is
\[
w(s,b)\equiv |1-iU(s,b)|^{-2}.
\]
 In the above formulas the variable $y$ denotes rapidity, i.e. $y=\sinh^{-1}(p/m)$,
where $p$ is a longitudinal momentum.
Thus, we can see that unitarity corrections are mostly important
at small impact parameters, i.e. they modify anisotropic flows at small centralities,
while peripheral collisions are almost not affected by unitarity.
The following limiting behavior of $v_n$ at $b=0$ can  easily be obtained:
\[
v_n(s,b=0, y,p_\perp)\to 0
\]
at $s\to\infty$ since $U(s,b=0)\to\infty$ in this limit.

In conclusion, one should note that inelastic events at the LHC energies come mainly from the 
non-zero values of the impact parameter. It is the result of approaching to unitarity saturation.

\section{Coherent rotation of transient matter and elliptic flow}
Before turning on discussion of anisotropic flows, it  is useful to note that the energy dependence of the average transverse momentum was considered in \cite{ptrans}.  
The mechanism is the following:  the nonzero orbital angular momentum  leads to coherent rotation
of a transient matter located in the overlap region as a whole  in the
$xz$-plane . This rotation is similar to the rotation of a liquid 
where strong correlations between particles momenta exist.

In this mechanism of hadron production, the valence constituent quarks excite a part of the cloud of virtual massive
quarks and those quarks  subsequently hadronize  and form the multiparticle final state. 

Here we use this mechanism
for evaluation of the elliptic flow in $pp$-interactions.
It was suggested in \cite{ptrans}  that the rotation of transient matter
will affect the average transverse momentum of secondary hadrons produced in proton-proton collisions.
A coherent rotation of transient liquid-like state results in a contribution to the transverse momentum
 in the form: 
\begin{equation} \label{ptl}
\Delta p_T=\kappa L(s,b),
\end{equation}
 where $L(s,b)$ is the orbital angular momentum  and $\kappa$ is a constant which has a dimension of inverse length.
Note that  the power-like dependence
of the average transverse momentum at high energies has been obtained in good agreement with experimental data \cite{ptrans}.

Going further, one should note that, in fact, the rotation gives a contribution to the $x$-component of the transverse 
momentum and does not contribute to the $y$-component of the transverse momentum, i.e.
\begin{equation} \label{ptx}
\Delta p_x=\kappa L(s,b)
\end{equation}
while
\begin{equation} \label{pty}
\Delta p_y=0.
\end{equation}
Thus, assuming that $p_x=p_0+\Delta p_x$ ($\Delta p_x \ll p_0$) and $p_y=p_0$, 
the effect of the rotating transient matter contribution  into the elliptic flow  can be calculated 
by analogy with the calculation of the average transverse momentum performed in \cite{ptrans}.
The resulting  integrated elliptic flow increases with energy
\[
v_2\propto s^{\delta_C} ,
\]
where  $\delta_C=0.207$. 
It should be noted here that transient matter consists of virtual constituent quarks strongly 
interacting
by pion (Goldstone bosons) exchanges .

The incoming constituent quark has a finite geometrical size determined by the radius $r_Q$ and 
interaction radius $R_Q$ ($R_Q>r_Q$). Former one is determined by the chiral symmetry 
breaking mechanism and the latter one --- by the confinement radius.
Meanwhile, it is natural to
suppose on the base of the uncertainty principle that size of the region where the virtual massive quark $Q$ is knocked out from the cloud
is determined by its transverse momentum, i.e. $\bar R\simeq 1/p_\perp$. However, it is
evident that $\bar R$ cannot be larger than the interaction radius of the valence
 constituent quark $R_Q$. 
It is also clear that $\bar R$ should not be less than the geometrical size of the valence constituent
quark $r_Q$ for this mechanism be a working one.  When  $\bar R$ becomes less than $r_Q$, 
this constituent quark mechanism does not work anymore
and one should expect vanishing collective effects in the relevant region of the transverse momentum.

The value of the quark interaction radius was
 obtained under analysis of the elastic scattering \cite{csn}
 and it has the following dependence on its mass
\begin{equation}\label{rq}
R_Q= \xi/m_Q \sim 1/m_\pi
\end{equation}
where $\xi \simeq 2$ and therefore $R_Q\simeq 1$ $fm$, while the geometrical radius of  quark $r_Q$
is about $0.2$ $fm$.
It should be noted  the region, which is responsible for the small-$p_\perp$ hadron production, has large transverse dimension and the incoming
constituent quark excites the rotating cloud of quarks with different values and directions
of their momenta in that case. Effect of rotation will therefore be smeared off  over the volume $V_{\bar R}$
 and then  one should expect that $\langle \Delta p_x \rangle_{V_{\bar R}} \simeq 0$. Thus,
\begin{equation}
\label{larg}
v^Q_2(p_\perp)\equiv\langle v_2\rangle_{V_{\bar R}}\simeq 0
\end{equation}
at small $p_\perp$.
When we proceed to the region of higher values of $p_\perp$, the radius $\bar R$ is decreasing
and the  effect of rotation  becomes more and more prominent, incoming valence quark excites now the region
where most of the quarks move coherently, in the same direction, with approximately
the same velocity. The mean value $\langle \Delta p_x \rangle_{V_{\bar R}} > 0$ and
\begin{equation}
\label{smal}
v^Q_2(p_\perp)\equiv\langle v^Q_2\rangle_{V_{\bar R}}> 0
\end{equation}
and it increases with $p_\perp$.
The increase of $v^Q_2$ with $p_\perp$ will disappear at
$\bar R =r_Q$, i.e. at $p_\perp \geq 1/r_Q$, and saturation will take place.
The value of transverse
momentum where the flattening starts is about $1$ $GeV/c$ for $r_Q\simeq 0.2$ $fm$. At very large transverse
momenta the constituent quark picture would not be valid and elliptic flow vanishes as it was already mentioned.

We discussed elliptic flow for the constituent quarks.
Predictions for the elliptic flow for the particular
hadrons depends on the supposed mechanism of hadronization. For the
region of the intermediate values of $p_\perp$ the constituent
quark coalescence mechanism \cite{volosh,volosh1} would be dominating one.
In that case the values for the hadron elliptic flow can be obtained from
the constituent quark one by the replacement $v_2\to n_Vv^Q_2$ and
$p_\perp\to p^Q_\perp/n_V$, where $n_V$ is the number of constituent quarks in the produced
hadron.

Typical qualitative dependence of elliptic flow in $pp$-collisions in this approach is presented in
Fig.1
\begin{figure}[h]
\begin{center}
  \resizebox{8cm}{!}{\includegraphics*{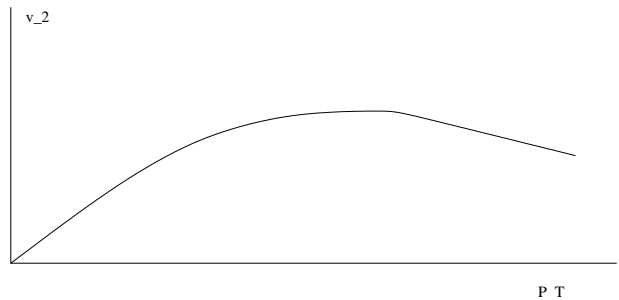}}
\end{center}
\caption{Qualitative dependence of the elliptic flow $v_2$ on transverse momentum in pp-collisions.}
\end{figure}

The centrality dependence of the elliptic flow is determined by dependence of 
the orbital angular momentum 
  $L$ on the impact parameter, i.e. it should   be decreasing towards  high and low centralities.
  Decrease toward high centralities is evident since no overlap of hadrons should occur at high enough
  impact parameters. Decrease of $v_2$ toward lower centralities is specific prediction of the proposed
  mechanism based on rotation
  since еру central collisions with smaller impact parameters would lead to slower rotation or its complete
   absence in the head-on collisions. Qualitative dependence of the elliptic flow on the impact parameter
  is similar to the curve depicted in Fig.1 where variable $b$ being used instead of transverse
 momentum.

 \section*{Conclusion}
 We considered the elliptic flow in the proton-proton interactions at the LHC energies in the particular nonperturbative
approach, where the origin of elliptic flow is associated with the effects of rotation of quark-pion transient matter.
This mechanism of anisotropic flow might be a leading one in hadron collisons, since those have smaller geometrical
extension and  the probabilty of hydrodynamical generation of elliptic flow is lower  compared to
the collisions of nuclei. 

At such high energies (LHC) the presence of reflective scattering mode serves as a trigger for
the mainly peripheral nature of the inelastic interactions and the presence of  large orbital angular 
momenta. Since correlations are maximal
in the rotation plane, a narrow ridge should be observed in  the two-particle correlation function \cite{ridge}. 

In conclusion, we would like to remark that 
performing   studies of the multiparticle productions
in $pp$--collisions at the LHC  under the scope of the
searches for the possible existence  of the rotation effects would be of a significant interest.
 Such effects would be absent when the genuine quark-gluon plasma
(gas of free quarks and gluons) being formed at the LHC energies because of all collective effects and 
the anisotropic flows, in particular, should dissapear in this case.

\end{document}